\documentclass[fleqn,usenatbib]{mnras}
\pdfoutput=1
\usepackage{amsmath}
\usepackage{txfonts}

\usepackage[T1]{fontenc}
\usepackage{ae,aecompl}

\usepackage{graphicx}	
\usepackage{caption}

\defcitealias{carlsten20a}{Carlsten et. al. (2020a)}
\defcitealias{carlsten20b}{Carlsten et. al. (2020b)}
\defcitealias{leaman20}{Leaman \& van de Ven (in prep.)}
\defcitealias{chies20}{Chies-Santos et. al. (in prep.)}
\defcitealias{denbrok14}{dB14}
\defcitealias{sanchez-janssen19}{SJ19}

\title[NSCs in faint Coma galaxies]{A high occurrence of nuclear star clusters in faint Coma galaxies, and the  roles of mass and environment}

\author[Zanatta et al.]{
Em\'ilio Zanatta$^{1}$\thanks{E-mail: emiliojbzanatta@ufrgs.br},
Rub\'en S\'anchez-Janssen$^{2}$,
Ana L. Chies-Santos$^{1}$,\newauthor
Rafael S. de Souza$^{3}$, 
John P. Blakeslee$^{4,5}$
\\
$^{1}$Departamento de Astronomia, Instituto de F\'isica, Universidade Federal do Rio Grande do Sul, Porto Alegre, R.S., Brazil\\
$^{2}$STFC UK Astronomy Technology Centre, Royal Observatory, Blackford Hill, Edinburgh, EH9 3HJ, UK\\
$^{3}$Key Laboratory for Research in Galaxies and Cosmology, Shanghai Astronomical Observatory,\\ Chinese Academy of Sciences, 80 Nandan Road, Shanghai 200030, China\\
$^{4}$Herzberg Astronomy and Astrophysics Research Centre, National Research Council of Canada, Victoria, BC, V9E 2E7, Canada\\
$^{5}$Gemini Observatory, NSF’s NOIRLab, Tucson, AZ 85719, USA
}

\date{Accepted XXX. Received YYY; in original form ZZZ}

\pubyear{2019}

\begin{document}
\label{firstpage}
\pagerange{\pageref{firstpage}--\pageref{lastpage}}
\maketitle

\begin{abstract}
We use deep high resolution  \textit{HST/ACS} imaging of two fields in the core of the Coma cluster to investigate the occurrence of nuclear star clusters (NSCs) in quiescent dwarf galaxies as faint as $M_{I} = -10$ mag. 
We employ a hierarchical Bayesian logistic regression framework to model the faint end of the nucleation fraction ($f_{n}$) as a function of both galaxy luminosity and environment. We find that $f_n$ is remarkably high in Coma: at $M_{I} \approx -13$ mag half of the cluster dwarfs still host prominent NSCs.
Comparison with dwarf systems in nearby clusters and groups shows that, within the uncertainties, the rate at which the probability of nucleation varies with galaxy luminosity is nearly universal. On the other hand, the fraction of nucleated galaxies at fixed luminosity does exhibit an environmental dependence. More massive environments feature higher nucleation fractions and fainter values of the half-nucleation luminosity, which roughly scales with host halo virial mass as $L_{I,f_{n50}} \propto \mathcal{M}_{200}^{-0.2}$. Our results reinforce the role of galaxy luminosity/mass as a major driver of the efficiency of NSC formation and also indicate a clear secondary dependence on the environment, hence paving the way to more refined theoretical models.
\end{abstract}




\section{Introduction}

At the central regions of galaxies of a wide range of masses,  luminosities,  and morphological types there exists a class of compact stellar systems known as nuclear star clusters (NSCs). These objects have half-light radii in the range of 1-50 pc, masses from $10^4$ to $10^8 \mathcal{M}_{\odot}$ and very extreme stellar densities comparable to  globular clusters (GCs) and ultra-compact dwarfs (UCDs)--but differ from these other compact systems in that they can exhibit a wide range of ages and metallicities. For a detailed overview of their properties we refer the reader to the comprehensive review by \citet{neumayer20}.

The formation of NSCs has been suggested to derive from two mechanisms, probably non-exclusive. One is a dissipationless process where already formed stellar clusters decay to the centre of the gravitational potential of the host galaxy and merge forming a large and dense structure \citep{tremaine75,arca-sedda14,gnedin14}. 
The other mechanism involves the inflow of gas to the central region of galaxies, where local star formation is triggered at higher rates than usual contributing substantially to their mass growth \citep{bekki03, bekki04, antonini13}. Not surprisingly, NSCs display a variety of scaling relations with their host galaxies involving both their stellar mass \citep{georgiev16, scott13} and stellar populations \citep{walcher05, turner12, georgiev14}. 
Finally, if GCs are suggested to contribute to the formation of NSCs, UCDs on the other hand might be the remains of disrupted nucleated galaxies \citep{drinkwater03a, pfeffer13,seth14a,neumayer20, ahn18, afanasiev18}. 

Amongst the most fundamental observables informing NSC formation scenarios are their occupation statistics. In other words, what galaxies host NSCs? 
Stellar nuclei occur almost across the entire spectrum of galaxy types. 
Historically, much of the early work on NSC demographics focused on galaxies in high density environments such as galaxy clusters, where number statistics are large. 
As a result, NSC occupation in early-type galaxies tends to be more robustly characterised than in late-types (but for details on star-forming hosts see \citealt{carollo98}, \citealt{georgiev09}, \citealt{georgiev14} and \citealt{neumayer20}). 
Recent studies have established that the nucleation fraction in quiescent galaxies exhibits a strong dependence on galaxy mass or luminosity, with a peak at $\sim$90\% around log($\mathcal{M}/\mathcal{M}_{\odot}) \approx 9$ followed by a steady decline toward both higher \citep{cote06,turner12,baldassare14a} and lower galaxy masses \citep{denbrok14,ordenes-briceno18a,sanchez-janssen19}.
Remarkably, while NSC occupation at the high-mass end seems to be rather universal, dwarf galaxies ($M_\odot \lesssim  10^9$) are now known to display a secondary dependence with the environment.
By comparing the nucleation fraction from the Next Generation Virgo Cluster Survey with literature data for other environments \citet{sanchez-janssen19} show that NSC occurrence is highest in Coma cluster dwarfs, followed by Virgo and Fornax, with the lowest nucleation fraction found in early-type satellites in the Local Group. 
This is consistent with early results showing that nucleated early-types tend to inhabit the inner, higher density regions of the Virgo cluster \citep{ferguson89,lisker07}. 
But \citet{sanchez-janssen19} expand on these studies to show that the behaviour holds at fixed galaxy mass, i.e, dwarfs of any given luminosity have a higher probability of being nucleated when they inhabit host haloes of larger virial masses. 
A shortcoming of that analysis is that the limiting magnitude of the Coma cluster sample \citep[][dB14 hereafter]{denbrok14} is significantly brighter than in all the other environments: $M_{I} \approx -13$ vs $M_{I} \approx -9$, well over an order of magnitude in luminosity.
As a result, the exact behaviour of the nucleation fraction in this most rich environment is not yet fully characterised--does it remain exceptionally high down to the faintest luminosities, or does nucleation become negligible for galaxies of comparable luminosity to those in the Virgo and Fornax cluster?
The present work aims at finally settling this question through the use of deep $HST/ACS$ imaging of the core of the Coma cluster to study the demographics of NSCs.
We also develop a novel Bayesian logistic regression framework to model the probability of nucleation, which enables us to self-consistently investigate its dependence on galaxy luminosity and environment for dwarfs in other clusters and groups.

This paper is structured as follows. In Section \ref{sec:data} we present the data used in this work, followed by a detailed description of the galaxy and NSC detection and measurement techniques in Section \ref{sec:phot}. In Section \ref{sec:logit} we present the statistical methodology developed to infer the nucleation fraction in Coma, which is compared to that of other environments in Section \ref{sec:results}. In Section \ref{sec:discussion} we discuss the main results within the context of observational and theoretical work on the galaxy nucleation fraction. Finally, in Section \ref{sec:conclusion} we summarise our results and their significance for NSC formation scenarios. Throughout this work we adopt a distance to the Coma cluster of $D=100$ Mpc \citep{carter08}, which corresponds to a physical scale of 485 pc/arcsec and a distance modulus of $(m-M) = 35$ mag.

\begin{figure*}
    \centering
    \includegraphics[scale=0.25]{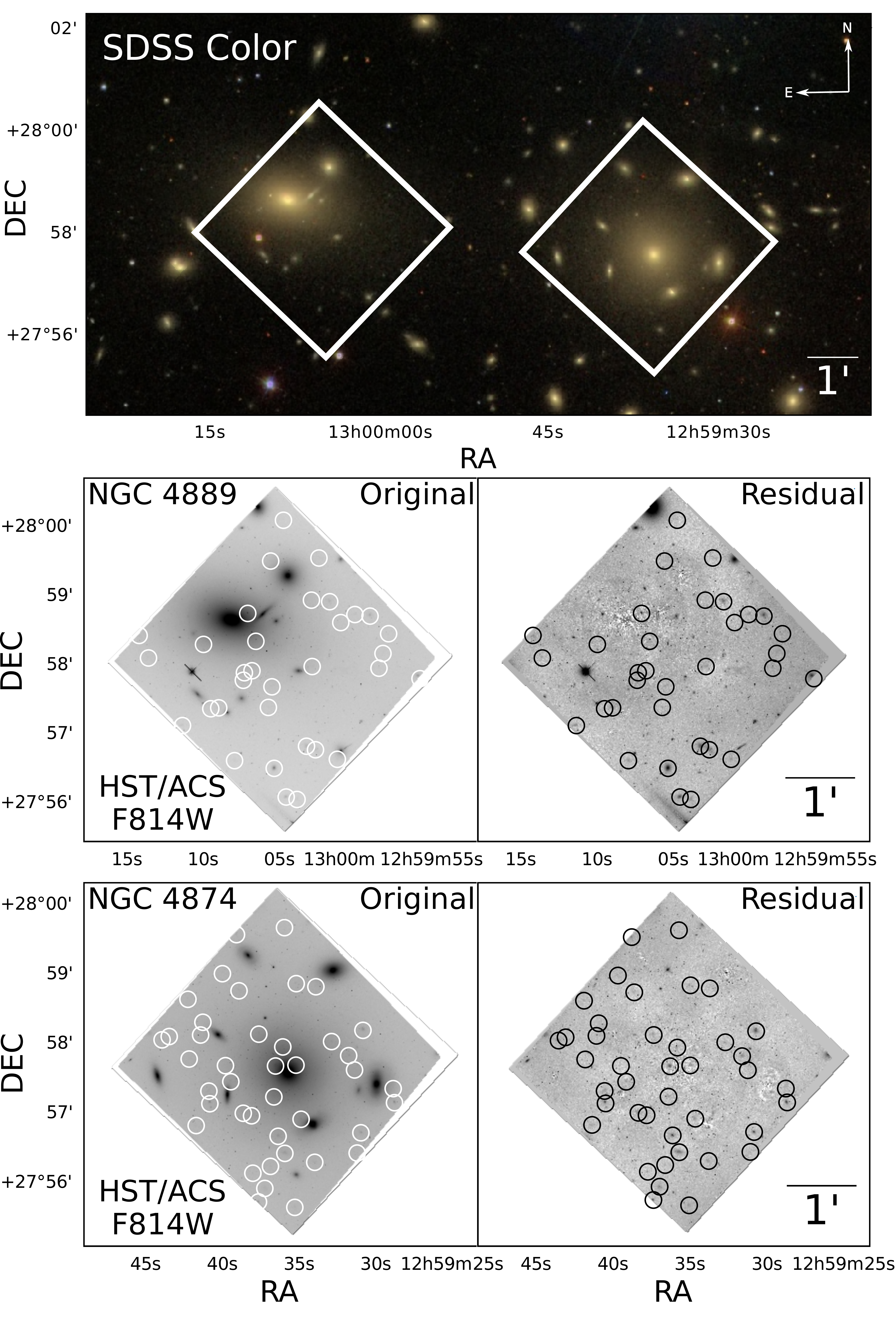}
    \caption{Top: SDSS colour-composite image of the central region of the Coma cluster, with white boxes representing the two HST/ACS pointings used in this work. Middle and bottom: The actual HST/ACS images used in this work, after \textsc{drizzle} treatment (left) and after the subtraction of bright galaxies to improve the detection of fainter objects (right). Circles indicate the positions of the detected galaxies in our sample (see Table \ref{tab:catalog}).}
    \label{fig:fields}
\end{figure*}

\section{Data}
\label{sec:data}

The Coma cluster data used in this work were obtained as part of Program GO-11711 (PI: J. Blakeslee) using the Advanced Camera for Surveys Wide Field Channel (ACS/WFC) onboard the Hubble Space Telescope (HST) in March 2012,  and consists of two fields centered on the bright cD galaxies NGC\,4874 and NGC\,4889. The observations run for four orbits with the F814W filter ($\approx$~$I$) and one orbit with the F475W filter ($\approx$~$g$). As a result, the former dataset is considerably deeper than the latter, and only the $I$ data are used throughout. The exposure times in $I$ are 10,425\,s and 9,960\,s for NGC\,4874 and NGC\,4889, respectively.

The data for NGC\,4874 were used previously in \citet{cho16} and we refer the reader to that work for additional details on the reduction steps, which are the same for the NGC\, 4889 field. Briefly, the images were dithered to fill the gap between the two ACS/WFC detectors, followed the standard pipeline processing from the STScI/Mikulski
Archive for Space Telescopes (MAST) and had the charge-transfer efficiency (CTE) correction algorithm of \citet{anderson10} applied. Finally, the CTE-corrected exposures were then processed with \textsc{apsis} \citep{blakeslee03} to produce the final corrected images shown in Fig. \ref{fig:fields}.

\begin{figure}
    \centering
    \includegraphics[width=\linewidth]{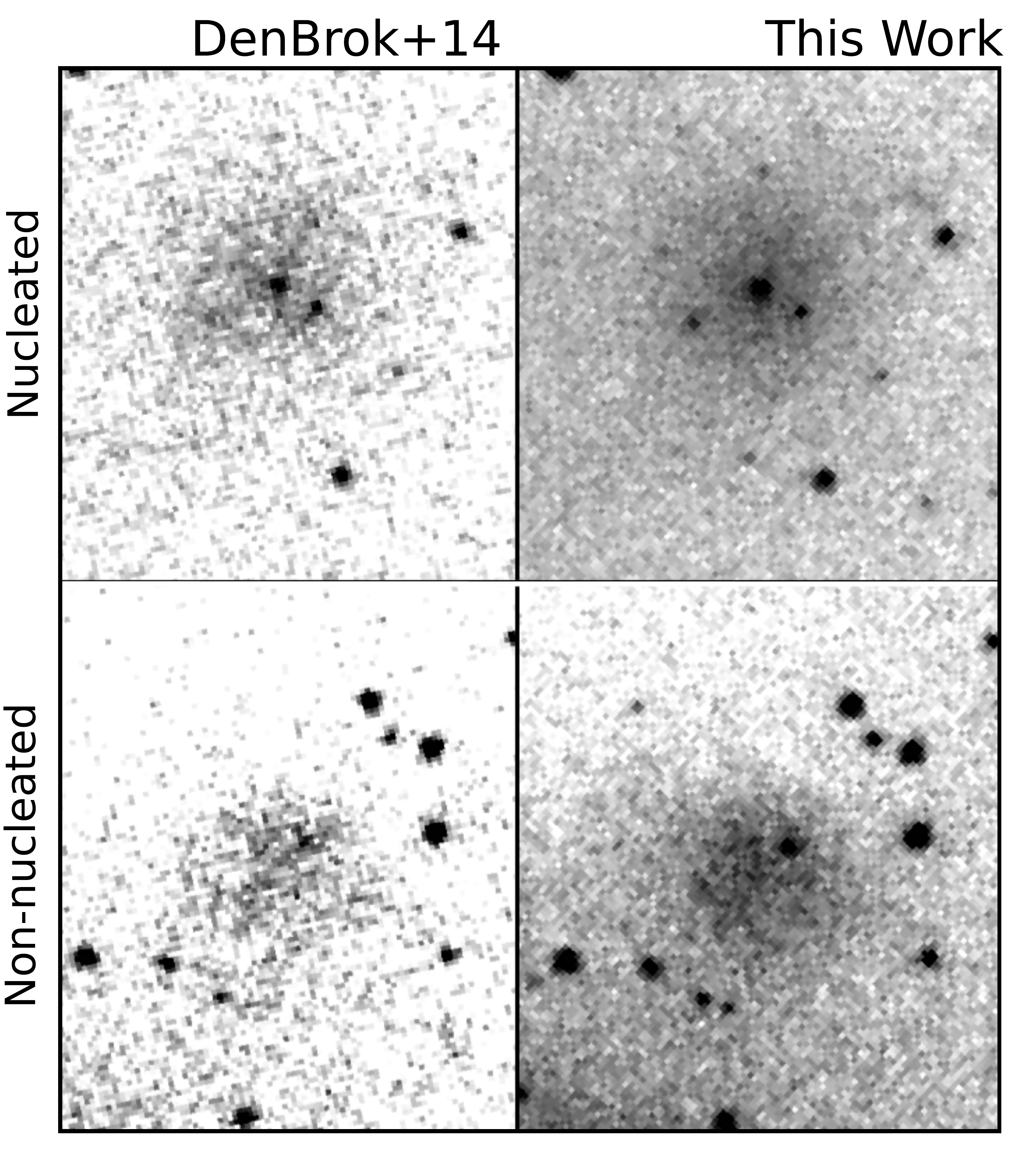}
    \caption{Comparison between the depth of images from \citetalias{denbrok14} and this work. On the top panels, two images of the same nucleated galaxy present in both the catalogue from \citetalias{denbrok14} and this work. On the bottom panel, a non-nucleated galaxy. All images are adjusted to the same scale and in the same HST/F814W filter.}
    \label{fig:denbrokcomp}
\end{figure}

The reference dataset for NSCs in the Coma cluster was introduced in \citetalias{denbrok14}, with imaging from the HST/ACS Coma Cluster Survey \citep{carter08}. 
The present study and that work are highly complementary. 
\citetalias{denbrok14} cover a large footprint and have robust number statistics at the bright end of the dwarf galaxy population ($M_{I} < -13$).
On the other hand, this work is limited to two ACS fields, but our deeper imaging - reaching $M_I \approx -10$ in comparison to the limiting magnitude of $M_I \approx -13$ of \citetalias{denbrok14} - allows us to probe much fainter galaxies and NSCs than ever before in Coma. 
In Fig.\ref{fig:denbrokcomp} we present a comparison between the $I$-band images used in this work and those from \citetalias{denbrok14}. The top row corresponds to a nucleated galaxy, whereas the bottom one shows a non-nucleated dwarf. 
The higher signal-to-noise ratio in our frames significantly improves on the detection and characterisation of NSCs and, especially, their low surface brightness hosts.

\subsection{NSCs in other environments from the literature}
\label{sec:other}
In addition to the Coma cluster, in this work we also analyse data for dwarf quiescent galaxies in the Virgo cluster \citep[from][]{sanchez-janssen19}, the Fornax cluster \citep[from][]{munoz15} and a collection of data for faint quiescent satellites in the local volume ($D < 12$ Mpc). The latter dataset is mainly drawn  from \citetalias{carlsten20a}, and further complemented with data from Chies-Santos et al. (in prep.) for dwarf companions of NGC\,3115. 
An important aspect of all these datasets is that they all feature roughly the same effective spatial resolution. 
This is a result of the superbly narrow point-spread function (PSF) delivered by HST which more than compensates for the much larger distance of the Coma cluster--and we therefore expect the NSC detection efficiency to be similar across the different environments. Moreover, the effective HST/ACS PSF FWHM in Coma is $\approx 37.92$ pc. NSCs in early-type dwarfs have typical sizes of $\lesssim 20$ pc \citep{cote06,turner12}, therefore at the distance of Coma they are all essentially unresolved in our images.   

As noted before, only early-type dwarfs are considered in the analysis. 
This is to avoid complications related to both the morphology-density relation as well as the notoriously difficult task of identifying NSCs in star-forming galaxies due to the presence of star formation and obscuration by dust.
More details on the literature data used in this work are presented in Appendix \ref{appendixA}.

With this choice of environments we are able to probe NSC occupation in host haloes with masses ranging from $5\times10^{15}$ in Coma to $10^{12} \mathcal{M}_{\odot}$ in the local volume.
When necessary, adopted mass estimates come from \citet{lokas03}, \citet{mclaughlin99} and \citet{drinkwater01} for Coma, Virgo and Fornax, respectively. 
For the local volume galaxies we derive a mean halo mass using the $log(V_{circ})-log(\mathcal{M}_{200})$ relation from the Illustris TNG100 simulations \citep{pillepich18} and the $V_{circ}$ values in Table 1 of \citetalias{carlsten20a}. Exceptions are: NGC 3115, for which the halo mass estimate comes from \citet{alabi17}; M81, from \citet{karachentsev02}; Cen A, from \citet{vandenbergh00}; M31, which comes from \citet{tamm12} and the Milky Way, from \citet{taylor16}.

\begin{figure*}
    \centering
        \includegraphics[width=\linewidth]{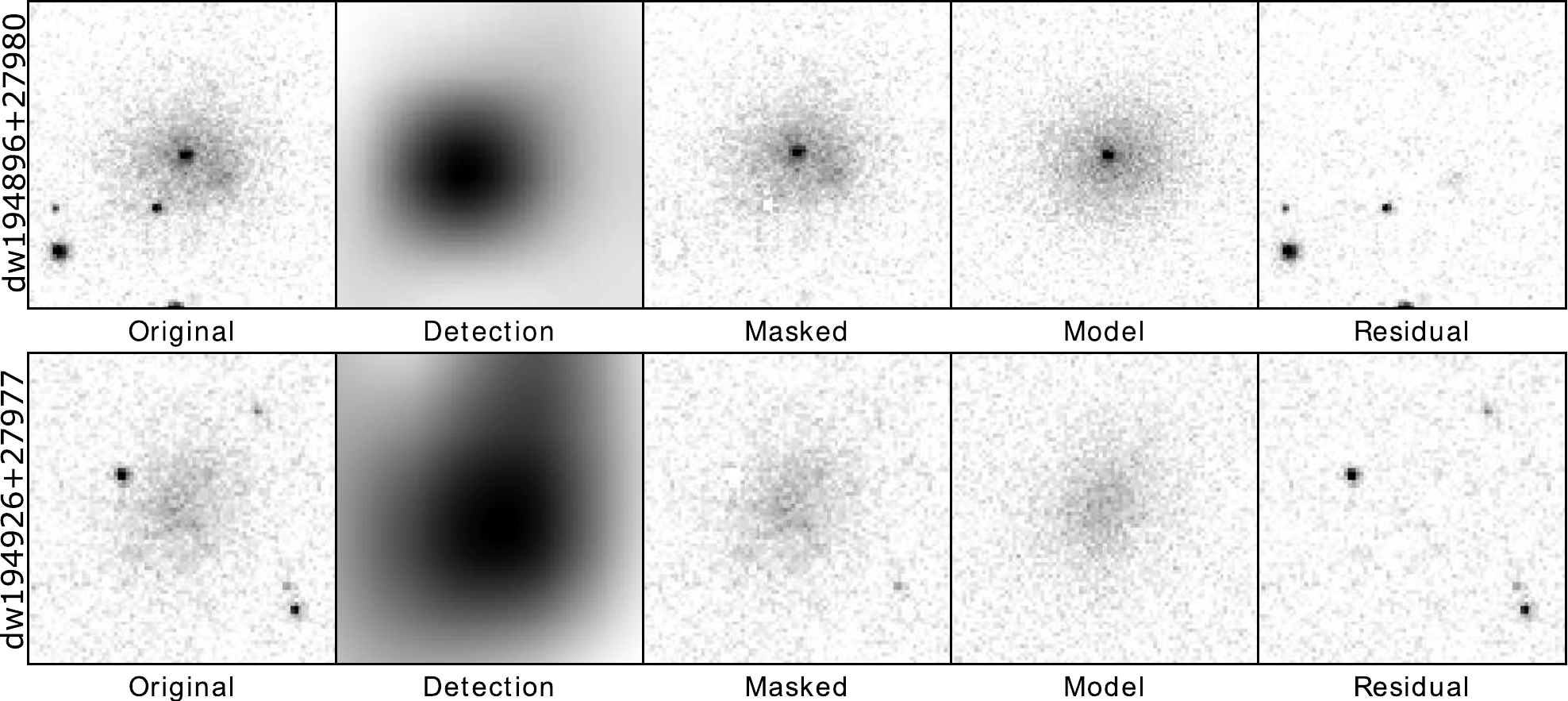}
    \caption{Summary of the procedure to detect and extract photometry for faint galaxies and their NSCs. From left to right: Section of the original image showcasing an example galaxy (nucleated on top, non-nucleated at the bottom). 
    In the second panel we show the \textsc{SourceExtractor} background image used for galaxy detection. Notice the significant increase in SNR,  which improves the detection limits. 
    The third panel contains the same image as the first panel, but now the point sources detected by the first \textsc{SourceExtractor} are masked--except for the central 6x6 pixels, which are unmasked to reveal the NSC.
    The fourth panel corresponds to the \textsc{galfit} model, to which we have added the typical noise of the $ACS$ images for representation purposes. 
    Finally, in the last panel we show the residual image from \textsc{galfit} modelling. Both galaxies are presented with the same scaling.
    }
    \label{fig:detectionpanel}
\end{figure*}

\section{Detection, membership and photometry}
\label{sec:phot}

In this work we build on the methods developed in recent surveys of the Virgo \citep{ferrarese16} and Fornax \citep{eigenthaler18} clusters to study the faint galaxy population and their star cluster systems.
Briefly, galaxy detection is carried out automatically using an algorithm optimised for the recovery of low surface brightness objects. 
Visual inspection of the candidates by one or more individuals follows, and cluster membership is assigned based on expected morphological features for early-type, quiescent dwarfs--namely, ellipsoidal shapes, smooth surface brightness profiles and absence of star formation features. 
Finally, photometric and structural parameters for the host galaxy and the NSC are derived through two-dimensional modelling of the galaxy images.
The detailed steps are as follows.

\subsection{Bright galaxy subtraction}

The two fields in this study are centred on  NGC\,4874 and NGC\,4889, the two dominant ellipticals in the core of the Coma cluster.  NGC\,4889 is the brightest galaxy in the cluster, but NGC\,4874 boasts an extended cD halo and resides somewhat closer to the centroid of the X-ray emission in the cluster.
The high density of bright satellite galaxies in these fields renders the detection of faint objects difficult. 
Therefore, the first step in our analysis consisted in the modelling and subtraction of the largest objects. 

We subtract the bright galaxies using the program \textsc{elliproof} \citep{elliproof,jordan04}, which fits a series of elliptical isophotes of varying centres, ellipticities, orientations, and low-order Fourier terms. The algorithm then interpolates smoothly between the isophotes and extrapolates outward beyond the last one. In both of our fields we model and subtract the brightest galaxy along with large neighbouring galaxies that affect the modelling of the central galaxy, as well as fainter galaxies that adversely affect the modelling of the bright satellite galaxies.  We adopt an iterative approach: subtracting the brighter galaxies, then subtracting the fainter neighbours, then remodelling the brighter galaxies with the neighbours subtracted, etc.  We iterate until we achieved a clean subtraction of all galaxies (about ten in each field) that were large enough to have a significant effect on the detection of the fainter objects we aim to study.

\subsection{\textsc{SourceExtractor} detection}

Galaxy detection is carried out with \textsc{SourceExtractor} \citep{bertin96} following a two-step approach. 
The first pass is optimised to extract point sources, whereas the second runs on the background image generated from the first one.
The process effectively acts as a low-pass spatial filter, resulting in a smoothed, high signal-to-noise ratio (SNR) image over scales larger than the PSF size.

In the first \textsc{SourceExtractor} run we set the minimum detection area, \textsc{detect\_minarea}, to 10 pixels above the detection threshold (\textsc{detect\_thres}) of 1.5$\sigma$ of the sky background. 
As described before, these parameters are set so as to detect compact objects such as GCs, foreground stars and background galaxies. 
We also set a background mesh size (\textsc{back\_size}) of 32 pixels with a 3x3 grid for the median filter (\textsc{back\_filtersize}). Using these parameters, \textsc{SourceExtractor} estimates the local
background in each mesh of rectangular grids across the entire image. In the second column of Fig. \ref{fig:detectionpanel} we show an example of such map, where the increase in SNR is evident.

We then proceed to the second run of \textsc{SourceExtractor}, this time on the smoothed image. We set \textsc{detect\_minarea} to 200 pixels above the \textsc{detect\_thres} of 1.2$\sigma$ of the sky background. 
We then match the positions of the detections with those in the original image, proceeding to visual inspection of the candidates to assign cluster membership and nucleation classification. 

\subsection{Visual classification}
Visual classification was independently carried out by two of the authors (EZ and RSJ).
Because faint early-type dwarfs have low surface brightness, the main contaminants are background late-type galaxies. 
We classify as members objects with smooth and spheroidal morphologies \citep{sanchez-janssen16}, and discard irregular galaxies or those displaying features consistent with ongoing star formation (clumps, arms, bars). 
We identify NSCs as compact spherical sources that in projection lie close to the geometric centre of the candidate galaxy (see Fig.\,\ref{fig:denbrokcomp}).
This requirement is later refined during the process of galaxy modelling (Sect.\,\ref{sec:modelling}).

We find that the two independent classifications fully agree down to a limiting magnitude of $m_{I} = 25$ mag.
Because the focus of this work is on the faint-end of the Coma galaxy population, we also set a bright limit of $m_{I} = 20$ mag.
The number of more luminous galaxies in the two fields under study is too low to be statistically meaningful. 
Our final catalogue is presented in Table \ref{tab:catalog}. We found 66 galaxies within $-15 < M_{I} < -10$ mag, 23 of which are first reported in this work. Among these, 33 have nuclear star clusters candidates.

\begin{table*}
\centering
\caption{Photometric and structural parameters for the galaxies detected in Coma obtained using \textsc{GALFIT} using \textsc{SourceExtractor} magnitude and positions as input parameters, as described in the text. From left to right: Identification for each galaxy, right ascension in degrees, declination in degrees, galaxy magnitude in the $I$ filter, NSC magnitude in the $I$ filter, S\'ersic index, effective radius in arcseconds, position angle in degrees, axis ratio and flags for previous detections of those galaxies. Such flags are as follows: 1: \citet{godwin83}; 2: \citet{iglesias-paramo03};
3: \citet{adami06a}; 4: \citet{yagi16}; 5: \citet{adami06b}; 6: \citet{hoyos11}; 7: \citetalias{denbrok14}. }
\label{tab:catalog}
\resizebox{\textwidth}{!}{%
\begin{tabular}{llllllllll}
\hline
ID & RA & DEC & $m_{I, gal}$ & $m_{I, NSC}$ & n & $R_{e}$ & b/a & PA & Prev. Detec. \\
& (deg) & (deg) & (mag) & (mag) & & (arcsec) &  & (deg) & \\
\hline
dw195019+27934  &  195.0191  &  27.9344  &  20.18  &  24.05  &  0.91  & 1.543 &  0.72  &  61.26   &  3,6        \\
dw195007+27981  &  195.0073  &  27.9815  &  20.20  &  23.34  &  1.49  & 2.465 &  0.88  &  11.15   &  2          \\
dw194870+27952  &  194.8701  &  27.9521  &  20.27  &  27.55  &  0.98  & 1.127 &  0.65  &  -22.84  &  2,6        \\
dw195029+27978  &  195.0296  &  27.9787  &  20.28  &  23.16  &  1.27  & 0.806  &  0.64  &  3.31    &  \\
dw194902+27960  &  194.9024  &  27.9609  &  20.40  &  25.11  &  0.74  & 1.210 &  0.80  &  63.29   &  \\
dw194920+27952  &  194.9202  &  27.9521  &  20.53  &  23.69  &  1.49  & 0.458  &  0.94  &  -38.14  &  2,6        \\
dw194905+27931  &  194.9052  &  27.9315  &  20.61  &  25.92  &  0.79  & 2.320 &  0.93  &  1.08    &  2,3,4      \\
dw194911+27949  &  194.9111  &  27.9496  &  20.66  &  23.68  &  1.49  & 2.976 &  0.78  &  43.22   &  3          \\
dw194879+27944  &  194.8791  &  27.9449  &  20.71  &  25.03  &  0.89  & 1.280 &  0.77  &  15.57   &  3          \\
dw195011+27945  &  195.0112  &  27.9459  &  20.78  &  25.15  &  0.61  & 2.028 &  0.68  &  4.02    &  3,6,7      \\
dw194907+27928  &  194.9070  &  27.9283  &  20.82  &  25.98  &  0.91  & 0.788  &  0.84  &  -57.99  &  2,3        \\
dw194882+27963  &  194.8823  &  27.9634  &  20.86  &  26.38  &  1.34  & 1.974 &  0.93  &  58.48   &  3,6,7      \\
dw194933+27967  &  194.9332  &  27.9672  &  20.93  &  26.24  &  0.96  & 1.728 &  0.63  &  29.73   &  3,6,7      \\
dw194923+27946  &  194.9239  &  27.9466  &  21.03  &  25.52  &  0.88  & 3.690 &  0.59  &  39.13   &  2,6        \\
dw194896+27961  &  194.8968  &  27.9612  &  21.05  &  26.09  &  1.01  & 1.887 &  0.81  &  85.96   &  \\
dw194895+27948  &  194.8954  &  27.9481  &  21.19  &  25.13  &  0.69  & 1.843 &  0.76  &  78.93   &  6          \\
dw194913+27992  &  194.9130  &  27.9925  &  21.19  &         &  0.82  & 1.728 &  0.55  &  -36.31  &  6,7        \\
dw194902+27953  &  194.9026  &  27.9535  &  21.33  &  24.78  &  1.32  & 2.289 &  0.75  &  24.00    &  6          \\
dw194870+27955  &  194.8704  &  27.9556  &  21.44  &  26.22  &  0.58  & 4.014 &  0.29  &  -55.01  &  \\
dw194920+27954  &  194.9204  &  27.9549  &  21.47  &  26.92  &  0.50  & 2.372 &  0.40  &  -86.76  &  2,6        \\
dw195037+27955  &  195.0375  &  27.9559  &  21.53  &  26.71  &  0.86  & 1.262 &  0.75  &  37.37   &  \\
dw194908+27949  &  194.9088  &  27.9490  &  21.65  &         &  0.83  & 0.949  &  0.89  &  -77.68  &  3,6,7      \\
dw195016+27933  &  195.0161  &  27.9338  &  21.69  &         &  0.52  & 1.655 &  0.67  &  -83.23  &  3,4,5,6,7  \\
dw194891+27937  &  194.8917  &  27.9378  &  21.75  &  26.70  &  0.63  & 1.563 &  0.77  &  16.76   &  2,3,5      \\
dw194900+27965  &  194.9004  &  27.9655  &  21.77  &  24.87  &  0.91  & 2.329 &  0.90  &  65.86   &  6          \\
dw194912+27979  &  194.9123  &  27.9790  &  21.77  &  25.79  &  0.65  & 1.226 &  0.68  &  3.71    &  6          \\
dw195000+27978  &  195.0003  &  27.9784  &  21.82  &  25.93  &  0.72  & 1.340 &  0.87  &  64.54   &  3,6,7      \\
dw195027+27971  &  195.0272  &  27.9719  &  21.82  &         &  0.69  & 2.759 &  0.75  &  66.77   &  \\
dw195047+27951  &  195.0475  &  27.9516  &  21.85  &         &  0.58  & 1.532 &  0.56  &  52.30    &  3,5        \\
dw194906+27968  &  194.9069  &  27.9686  &  21.86  &         &  0.85  & 1.569 &  0.59  &  -31.93  &  6          \\
dw195033+27943  &  195.0332  &  27.9432  &  21.89  &  26.56  &  0.79  & 1.456 &  0.76  &  40.55   &  3,5,6,7    \\
dw195030+27964  &  195.0306  &  27.9644  &  21.90  &         &  0.88  & 1.493 &  0.75  &  86.83   &  3          \\
dw194897+27927  &  194.8971  &  27.9270  &  21.97  &  26.50  &  0.83  & 1.277 &  0.90  &  -64.46  &  3,5        \\
dw195058+27973  &  195.0589  &  27.9731  &  22.03  &  26.10  &  0.70  & 1.364 &  0.81  &  24.79   &  3          \\
dw195010+27992  &  195.0100  &  27.9920  &  22.05  &         &  0.77  & 2.559 &  0.57  &  11.08   &  \\
dw194896+27980  &  194.8969  &  27.9806  &  22.10  &  26.38  &  0.71  & 0.996  &  0.83  &  80.16   &  3,6        \\
dw194887+27966  &  194.8871  &  27.9667  &  22.20  &  26.15  &  0.93  & 1.124 &  0.81  &  27.66   &  3,6,7      \\
dw194992+27969  &  194.9925  &  27.9692  &  22.35  &         &  0.92  & 1.496 &  0.99  &  -65.99  &  3,6        \\ \hline

\end{tabular}%
}
\end{table*}

\begin{table*}
\centering
\ContinuedFloat
\caption{Photometric and structural parameters for the galaxies detected in Coma. (continued) }
\resizebox{\textwidth}{!}{%
\begin{tabular}{llllllllll}
\hline
ID & RA & DEC & $m_{gal}$ & $m_{NSC}$ & n & $R_{e}$ & PA & b/a & Prev. Detec. \\
& (deg) & (deg) & (mag) & (mag) & & (arcsec) & (deg) & & \\
\hline
dw194914+27957  &  194.9144  &  27.9573  &  22.49  &         &  0.51  & 0.813 &  0.79  &  30.38   &  6          \\
dw195005+27943  &  195.0052  &  27.9435  &  22.51  &         &  0.76  & 1.060 &  0.62  &  -65.89  &  \\      
dw195011+27965  &  195.0117  &  27.9658  &  22.58  &  25.23  &  0.99  & 1.354 &  0.75  &  42.23   &  \\
dw195022+27961  &  195.0229  &  27.9611  &  22.66  &         &  0.99  & 1.743 &  0.57  &  85.86   &  \\
dw194993+27965  &  194.9939  &  27.9655  &  22.68  &  23.68  &  1.49  & 0.811 &  0.71  &  35.60   &  \\
dw195023+27991  &  195.0231  &  27.9913  &  22.72  &  25.42  &  0.53  & 2.215 &  0.63  &  37.20    &  \\
dw194908+27935  &  194.9086  &  27.9351  &  22.84  &         &  0.56  & 0.720 &  0.84  &  -57.49  &  3          \\
dw195028+27964  &  195.0285  &  27.9649  &  22.84  &         &  0.50  & 0.996 &  0.94  &  -52.86  &  \\
dw195030+27962  &  195.0308  &  27.9626  &  22.90  &         &  0.94  & 1.518 &  0.65  &  -24.14  &  3          \\
dw194926+27977  &  194.9261  &  27.9770  &  23.29  &         &  0.66  & 0.802 &  0.77  &  32.00    &  3,6        \\
dw194931+27968  &  194.9312  &  27.9680  &  23.29  &         &  0.50  & 1.354 &  0.71  &  50.28   &  \\
dw195024+27956  &  195.0240  &  27.9560  &  23.33  &         &  0.75  & 1.187 &  0.59  &  -8.06   &  3          \\
dw195056+27968  &  195.0567  &  27.9680  &  23.33  &         &  0.93  & 1.059 &  0.73  &  -32.45  &  3          \\
dw194916+27961  &  194.9161  &  27.9610  &  23.41  &         &  0.67  & 0.965 &  0.74  &  -46.73  &  6          \\
dw194903+27936  &  194.9037  &  27.9368  &  23.48  &         &  0.52  & 0.737 &  0.90  &  54.27   &  3          \\
dw194891+27979  &  194.8915  &  27.9799  &  23.57  &         &  0.84  & 0.674 &  0.71  &  26.25   &  6          \\
dw195019+28001  &  195.0199  &  28.0011  &  23.59  &         &  0.62  & 0.604 &  0.80  &  1.25    &  3          \\
dw195039+27955  &  195.0397  &  27.9557  &  23.67  &         &  0.67  & 0.838 &  0.66  &  -9.15   &  \\
dw194880+27940  &  194.8802  &  27.9400  &  23.68  &         &  0.68  & 0.747 &  0.69  &  7.66    &  3          \\
dw194880+27960  &  194.8808  &  27.9601  &  23.79  &         &  1.12  & 0.798 &  0.76  &  -27.25  &  \\
dw194916+27983  &  194.9167  &  27.9831  &  24.08  &         &  0.61  & 0.724 &  0.71  &  28.21   &  \\
dw194990+27973  &  194.9909  &  27.9739  &  24.25  &         &  0.99  & 0.854 &  0.85  &  47.79   &  \\
dw194922+27971  &  194.9221  &  27.9714  &  24.40  &         &  0.50  & 0.670 &  0.76  &  75.78   &  \\
dw195012+27982  &  195.0121  &  27.9820  &  24.44  &         &  0.59  & 0.784 &  0.65  &  40.80    &  \\
dw194922+27968  &  194.9225  &  27.9684  &  24.45  &         &  0.85  & 0.405 &  0.63  &  -63.66  &  \\
dw194925+27962  &  194.9257  &  27.9626  &  24.56  &         &  0.65  & 0.511 &  0.72  &  -48.45  &  3,6        \\
dw195041+27971  &  195.0417  &  27.9712  &  24.81  &  27.74  &  0.56  & 0.675 &  0.75  &  33.48   &  \\
dw195004+27976  &  195.0041  &  27.9766  &  24.90  &         &  0.50  & 0.502 &  0.66  &  -55.04  &  \\ \hline
\end{tabular}%
}
\end{table*}
\subsection{GALFIT modelling}
\label{sec:modelling}

To determine the structural and photometric properties of the candidate galaxies and their NSCs we model their surface brightness profiles using \textsc{galfit} \citep{peng02}.
We use the segmentation maps from the first \textsc{SourceExtractor} run to mask all objects around our detected galaxies, except for the central point sources in the visually-identified nucleated dwarfs. Furthermore, we use PSF models obtained with \textsc{PSFex} \citep{bertin11}.
We use as initial conditions the \textsc{mag\_auto} magnitudes and \textsc{flux\_radius} results from the \textsc{SourceExtractor} catalogue, as well as an initial S\'ersic index of $n$=0.75, position angle of 45 degrees and axis ratio of 0.8. The non-nucleated galaxies are modelled with a single S\'ersic profile, while for the nucleated ones we use a S\'ersic profile alongside a PSF component. 

Initial magnitudes for the PSF component are the \textsc{mag\_auto} measured by the first \textsc{SourceExtractor} run. 
The two components have initial central positions defined by the positions detected by \textsc{SourceExtractor} for the galaxy or the NSC, in the case of nucleated objects. 

To aid in the modelling of the fainter galaxies we employ constraints to the parameters to be fitted by \textsc{galfit}. Based on the structural parameters for faint galaxies in the Virgo cluster \citep{ferrarese20} we limit the S\'ersic index to vary in the range $0.5 \leq n \leq 1.5$. Constraining the S\'ersic index aids \textsc{galfit} in converging to realistic values for other structural parameters even for faint, low surface brightness galaxies. 
We also constrain the relative position of the S\'ersic and PSF components to be within three pixels of each other. 
This is inspired by $HST$ studies of NSCs showing that stellar nuclei are rarely offset from the geometric centre of the host galaxy \citep{cote03,turner12}.
It also guarantees that the likelihood of contamination from chance projection of stars or GCs over the galactic body remains insignificant, with the mean number of such contaminants estimated to be $\approx$ 0.002 per galaxy. This is calculated by counting the total number of point sources (\textsc{class\_star} $ \geq 0.6$) that have a magnitude difference of less than 0.5 mag with respect to each NSC. The surface density of such candidates is then multiplied by the area enclosed in a circle with a radius of three pixels, which is our criterion for a bona-fide NSC detection.
Results are not sensitive to the magnitude difference between putative contaminants and the candidate NSCs.

\begin{figure}
    \centering
    \includegraphics[width=\linewidth]{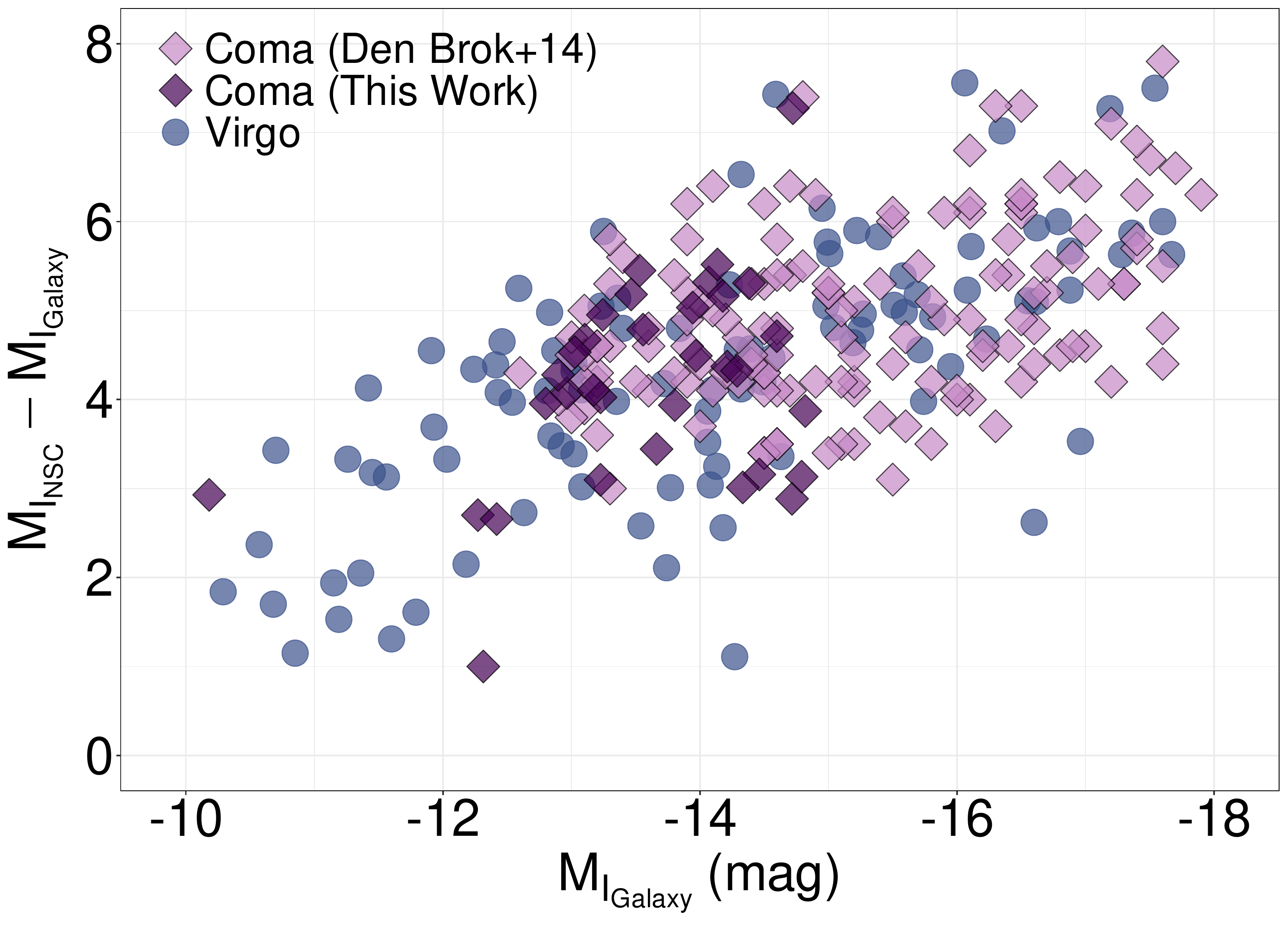}
    \caption{Difference between the magnitude of the nuclei, $M_{I, NSC}$ and the one of its host galaxy, $M_{I, Galaxy}$, for all nucleated galaxies in the Coma cluster sample from this work (shown in table \ref{tab:catalog}, purple diamonds) and \citetalias{denbrok14} (magenta diamonds), as well as nucleated galaxies in the Virgo cluster (from \citet{sanchez-janssen19}, blue circles), as a function of the host galaxy absolute magnitude. For both environments, brighter galaxies tend to show larger differences in magnitude from their nuclei, although a scatter is also evident, showcasing its stochastic nature.}   \label{fig:nucgal}
\end{figure}

A summary of the method is shown in Fig.\,\ref{fig:detectionpanel}. The galaxy properties obtained from this procedure are presented in Table \ref{tab:catalog}, where magnitudes are dereddened from Galactic extinction using the maps from \citet{schlafly11}.
We find that our catalogue has eight common entries with that from \citetalias{denbrok14}. The root mean square deviation of the galaxy magnitudes is only 0.12 mag. See Appendix \ref{appendixC} for a full description of our photometric uncertainty estimation. This was done by applying our photometry procedure to 10,000 mock galaxies randomly positioned in the two HST fields used in this work. Based on the results of these simulations, we estimate mean uncertainties of $\delta m_{I, gal} \sim 0.2$ mag for the magnitudes obtained in Table \ref{tab:catalog}. Uncertainties to the other parameters are shown in Appendix \ref{appendixC}.

In Fig. \ref{fig:nucgal} we show the difference between the magnitude of the NSC and its host galaxy for all the nucleated galaxies in our sample, as well as those from \citetalias{denbrok14} and the nucleated galaxies in the Virgo cluster from \citet{sanchez-janssen19}. Similarly to what was found in the two latter works, the relative contribution of the NSC to the overall brightness of the host galaxy decreases with galaxy luminosity--albeit with a large scatter in the relation. 
From the perspective of comparing different environments, we see that the trend in Fig. \ref{fig:nucgal} is very similar for both Virgo and Coma galaxies.
Overall, it seems clear that the luminosity of the NSC is related to that of the host galaxy regardless of the environment, but the large scatter in the relation indicates that NSC growth varies substantially from one galaxy to another.

\section{Statistical Modelling of the nucleation fraction}
\label{sec:logit}

To analyse the nucleation fraction in the Coma cluster and how it compares  to other environments  we employ Bayesian logistic regression \citep[see e.g.][for a detailed description]{Hilbe2017}.  
Logistic regression belongs to the family of generalised linear models, 
and is particularly suitable  for handling Bernoulli-distributed (binary) data. Such distribution characterises processes with two possible outcomes $\{0,1\}$, be it success or failure, yes or no, or alike -- in our case it is nucleation or non-nucleation.
Previous applications of logistic models in Astronomy include the studies of star-formation activity in primordial dark matter halos \citep{deSouza2015AeC}, the escape of ionising radiation at high-redshift \citep{Hattab2019},  the effect of environment in the prevalence of Seyfert galaxies \citep{desouza16}, and the redshift evolution of UV upturn galaxies \citep{Dantas2020}.
While the full behaviour of the nucleation fraction departs significantly from the logistic relation \citep{sanchez-janssen19}, we make use of the fact that for all studied environments $f_n$ appears to peak at masses log($\mathcal{M}/\mathcal{M}_{\odot}$) $\approx$ 9 \citep{neumayer20}, and then declines gradually toward lower masses until it becomes negligible.  

The regression model is the following:
\begin{align}
\label{eq:model}
   &\rm y_{i}\sim \rm {Bern}\left(p_{i}\right), \notag   \\
   &\eta_{i} \equiv \log\left(\frac{p_{i}}{1-p_{i}}\right),\\
    & \rm \eta_{i} =  \beta_{1[k]}+\beta_{2[k]}\,M_{I,i},\notag \\
   & \begin{bmatrix}\notag
\beta_{1[k]}     \\
\beta_{2[k]}     
    \end{bmatrix}
    \sim \rm {Norm}{\left(   \begin{bmatrix}\notag
\mu_{\beta}     \\
\mu_{\beta}     
    \end{bmatrix},\Sigma\right); \quad \Sigma \equiv 
    \begin{bmatrix}\notag
\sigma_{\beta}^2   & 0  \\
0  & \sigma_{\beta}^2  
    \end{bmatrix}},\\
    &\mu_{\beta} \sim \rm{Norm(0,10^2)}; \quad \notag
   \sigma_{\beta}^2 \sim \rm{Gamma(0.1,0.1)}.
    \end{align}
The above model reads as follows. 
Each of the $i$-th  galaxies in the dataset has its probability to manifest nucleation modelled as a Bernoulli process, whose probability of success relates to $M_{I,i}$ through a logit link function, $\eta_{i}$ (to ensure the probabilities will fall between 0 and 1),  where the index $k$ encodes the cluster/group environment. A subtle, but important characteristic of our model is the treatment of the intercept $\beta_{1[k]}$ and slope $\beta_{2[k]}$ coefficients via hierarchical partial pooling. For the case studied here, it falls under the umbrella of generalised linear mixed models \citep[see e.g.][for details]{Hilbe2017}. A simple intuition behind this choice is given below. When modelling the same relationship across multiple groups, there are three common  choices: pooled, unpooled, and partially-pooled models. In our case the pooled model implies a fit to the entire data, completely ignoring potential differences across cluster/groups. In other words, this would implicitly assume a universal shape for the nucleation fraction. In the other extreme lies the ubiquitous unpooled model, which implies fitting each individual case,  ignoring any potential correlation  across cluster/groups. While it seems a harmless choice, this model is very sensitive to differences in sample size and magnitude range between different cluster/group environments. The  most conservative option is partial pooling, which infers different parameters for each group, but  allows them to share information. This is done via the use of hyper-priors for its coefficients.  This is included in our model by assuming  a multi-Normal prior for $\beta_{1[k]}$ and $\beta_{2[k]}$ with a common mean $\mu_{\beta}$ and variance $\sigma_{\beta}^2$, to which we assigned  weakly informative Normal and Gamma hyper-priors respectively. 

We evaluate the model using the Just Another Gibbs Sampler (\textsc{jags}\footnote{http://cran.r-project.org/package=rjags}) package  within the \textsc{R} language \citep{rcore19}. 
We initiate three Markov Chains by starting the Gibbs sampler at different initial values sampled from a Normal distribution with zero mean and standard deviation of 100. Initial burn-in phases were set to 5,000 steps followed by 20,000 integration steps, which are sufficient to guarantee the convergence of each chain, following Gelman-Rubin statistics \citep{gelman-rubin92}.

\begin{table}
\centering
\caption{Summary of the parameters estimated from the model presented in eq. \ref{eq:model}. In the first column we show $M_{I,f_{n50}}$, the magnitude at which the estimated probability of nucleation reaches 50\%. In the second column $\Delta$Odds represents the expected change in the odds of nucleation by a variation of one unit of magnitude. In the last two columns, $\beta_1$ and $\beta_2$ are the mean posteriors for the intercept and slope, respectively, of the link function $\eta_{i}$. One can see that $\Delta$Odds is within a 10\% difference among all environments, while the variation of $M_{I,f_{n50}}$ present a significant difference in luminosity.}

\label{tab:model}
\resizebox{\linewidth}{!}{%
\begin{tabular}{lllll}
                                      & $M_{I,f_{n50}}$                                      & $\Delta$Odds & $\beta_{1} (intercept)$   & $\beta_{2} (slope)$    \\ \hline
\multicolumn{1}{l|}{Coma}             & $-12.98${\raisebox{0.5ex}{\small$_{-0.25}^{+0.26}$}} & -52.2\%      & $-9.66 \pm 1.56$          & $-0.74 \pm 0.11$ \\
\multicolumn{1}{l|}{Virgo}            & $-14.39${\raisebox{0.5ex}{\small$_{-0.25}^{+0.25}$}} & -50.0\%      & $-9.89 \pm 1.28$          & $-0.70 \pm 0.10$ \\
\multicolumn{1}{l|}{Fornax}           & $-14.16${\raisebox{0.5ex}{\small$_{-0.31}^{+0.31}$}} & -47.3\%      & $-9.25 \pm 0.98$          & $-0.64 \pm 0.07$ \\
\multicolumn{1}{l|}{Local Early-Type} & $-14.64${\raisebox{0.5ex}{\small$_{-0.89}^{+0.98}$}} & -41.4\%      & $-7.94 \pm 1.76$          & $-0.54 \pm 0.14$ \\
\multicolumn{1}{l|}{Local Late-Type}  & $-15.38${\raisebox{0.5ex}{\small$_{-0.89}^{+0.96}$}} & -47.5\%      & $-9.98 \pm 2.21$          & $-0.66 \pm 0.17$ \\ \cline{1-5}
\end{tabular}
}
\end{table}

\section{Results}
\label{sec:results}

\begin{figure}
    \centering
    \includegraphics[width=\linewidth]{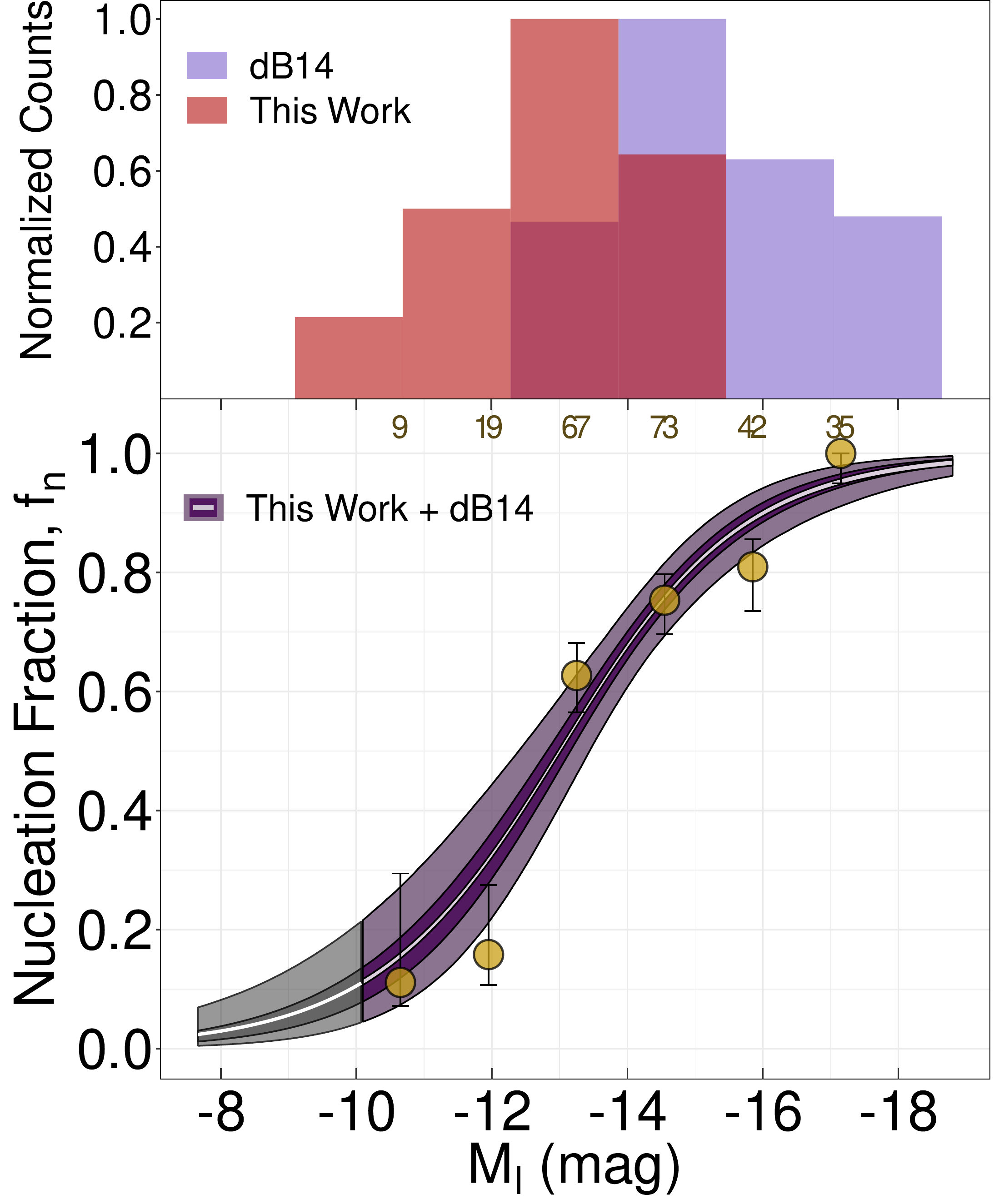}
    \caption{\textit{Top:} Distributions, in normalised counts, for the data obtained in this work and presented in Table \ref{tab:catalog} (red histogram) and the catalogue of Coma galaxies from \citetalias{denbrok14} (blue histogram). \textit{Bottom}:Nucleation fraction versus absolute magnitude for galaxies in Coma, combining the data from this work and the one from \citetalias{denbrok14}. In the cases where galaxies were detected in the two datasets, we keep the magnitudes from our own analysis. The white curve is the mean posterior from the Bayesian logistic regression. The purple shaded regions show the 50\% and 95\% confidence intervals, whereas the grey shades indicate the magnitudes where the model extrapolates the data. The yellow solid circles represent the median nucleation fraction in a binned representation of the data, with uncertainties given by the corresponding 68\% Bayesian credible intervals. The number of objects in each bin is shown at the top.}   \label{fig:nucl}
\end{figure}

\begin{figure*}
    \centering
    \includegraphics[width=\linewidth]{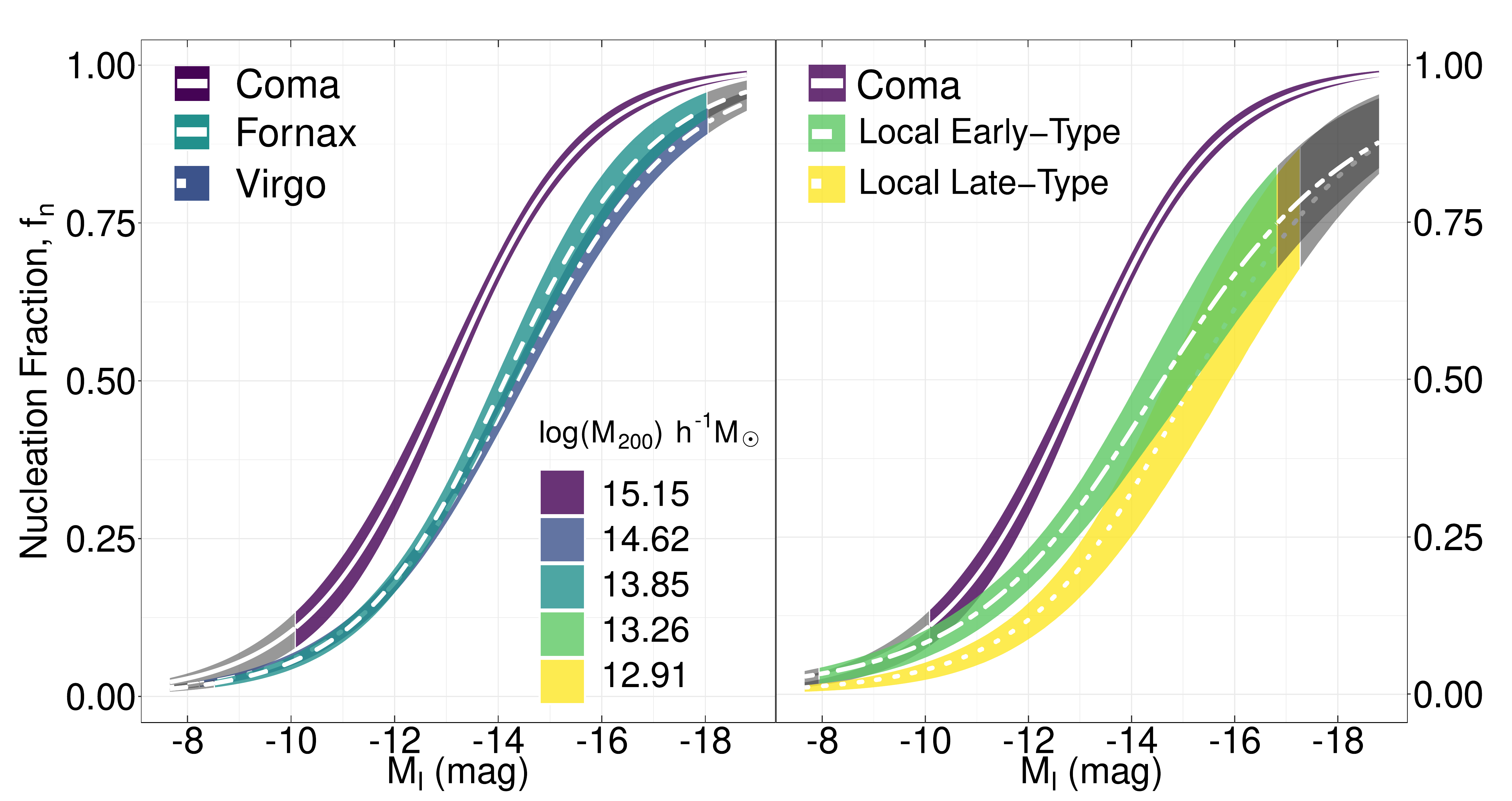}
    \caption{Mean posteriors from the Bayesian logistic regression models and 50\% confidence intervals, in the left panel, for the Coma cluster data from this work and \citetalias{denbrok14}, Virgo cluster from \citet{sanchez-janssen19} and Fornax cluster from \citet{munoz15}. In the right panel, the Coma cluster again as well as a combined sample of galaxies in the local volume in groups with a central Early or Late-Type galaxy (see appendix \ref{appendixA} for references). The samples are colourised by the approximated cluster/group virial halo mass, from the references mentioned in the text. For the case of the local volume environments, the halo mass presented is a mean of the halo masses of the sample galaxies within it. The grey regions of the curves represent the regions where the model extrapolates the actual data. Coma galaxies show a higher probability of nucleation than all other environments at all magnitudes considered.}   \label{fig:nuclboth}
\end{figure*}

When modelling the nucleation fraction the Coma sample presented here is merged with the catalogue from \citetalias{denbrok14} for completeness purposes at the bright end. The data from this work comprises Coma galaxies with magnitudes in the $-10.0 < M_{I} < -15.0$ range, while \citetalias{denbrok14} data ranges from $-12.8 < M_{I} < -19.0$.
In the few cases where a galaxy is detected in the two studies we keep the magnitudes from our own analysis.

\subsection{Nucleation fraction in the Coma cluster}

Fig. \ref{fig:nucl} shows the Coma nucleation fraction, $f_n$, as a function of  galaxy absolute magnitude, $M_{I}$. The shaded areas indicate the 50\% and 95\% probability intervals around the mean of the posterior for the logistic model (white curve).
The means and 68\% confidence level values of the corresponding coefficients are shown in Table \ref{tab:model}. 
The grey region represents the extrapolated solution beyond which there is no data coverage.
For visualisation purposes we also show with yellow circles the nucleation fraction calculated in eight equal-sized bins. 
The corresponding error bars show the 68\% Bayesian confidence level. 
The tick marks at the top and bottom of the panel indicate the magnitudes for each of the nucleated and non-nucleated galaxies, respectively, and the yellow histogram is a binned representation with the purpose to highlight the prevalence of NSCs in brighter galaxies.

We confirm the strong dependence of the nucleation fraction on galaxy luminosity in Coma: it peaks at nearly $f_n \approx$ 100\% at $M_{I} = -18$ (the characteristic luminosities of classical dE galaxies) and then declines to become almost negligible at $M_{I} = -10$. 
Overall, the nucleation fraction is remarkably high, with more than half of the $M_{I} = -13$ dwarf galaxies still hosting NSCs.
This is clearly shown by both the binned nucleation fraction and the logistic model, but only the latter is able to produce a smooth solution while also allowing us to study the rate at which NSCs occur in galaxies of different brightness.
More importantly, we are now in a position to quantitatively compare the exact shape of the nucleation fraction in Coma and in other environments.

\subsection{Nucleation in Other Environments}

In Fig.\ref{fig:nuclboth} we show the mean posterior for the nucleation fraction in the Coma cluster compared to that in the Virgo and Fornax clusters (left panel), as well as in nearby groups (right panels).
In order to improve the number statistics in these less rich environments we subdivide and stack the samples of group dwarfs into satellites of early- and late-type centrals (see Appendix\,\ref{appendixA} for details). 
The total number of quiescent satellites in the subsamples of early- and late-type groups is 93 and 90, respectively.
As with Coma, the  means and 68\% confidence level values of the  coefficients for the logistic model are shown in Table \ref{tab:model}. 
Two results are readily apparent.

\begin{figure}
    \centering
    \includegraphics[width=\linewidth]{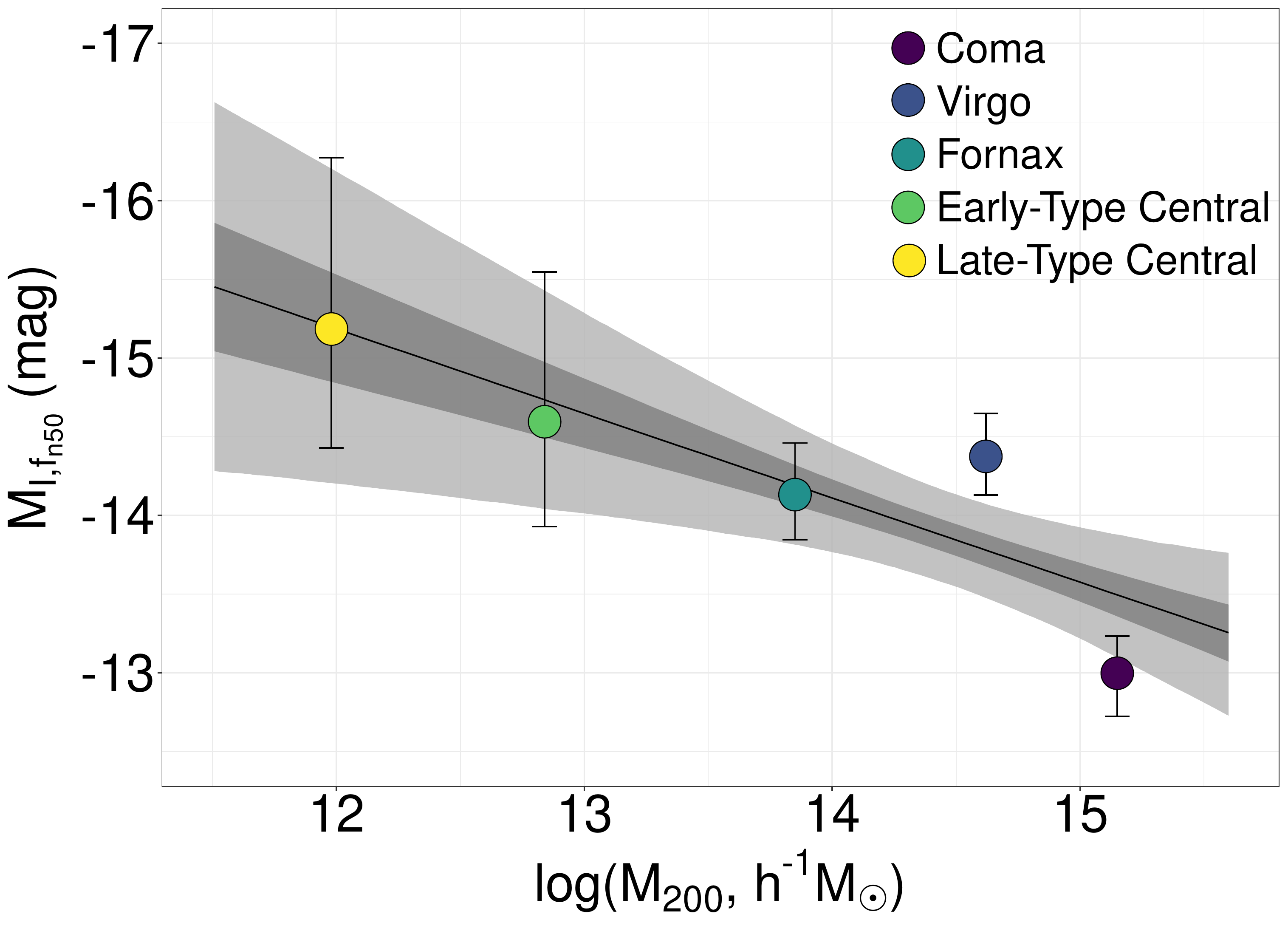}
    \caption{Mean posterior for the magnitude at which the probability of nucleation is 50\%, $M_{I,fn50}$, as a function of host virial halo mass (references mentioned in the text). The mean, 50\% and 95\% confidence intervals for a linear model are indicated by the solid line and grey shaded areas, respectively. More massive environments exhibit fainter values of the half-nucleation magnitude.}
    \label{fig:hostmass}
\end{figure}

\begin{itemize}
    \setlength\itemsep{1em}
    
    \item[1)] \emph{Nucleation has a nearly universal dependence on dwarf galaxy luminosity}.
    This is shown in Table\,\ref{tab:model} by the very similar $\beta_{2}$ slope parameters for all the different environments, which are consistent with each other at the 1\,$\sigma$ level. In the table we also show the corresponding $\Delta$Odds, which represents the change in nucleation odds given a variation of one unit in magnitude for each environment, i.e., the nucleation rate by magnitude. We find very similar values for all environments, perhaps with the exception of systems dominated by early-type centrals where the nucleation fraction displays a slightly shallower decline.
    
    \item[2)] \emph{Nucleation is more common in more massive haloes regardless of dwarf galaxy luminosity}. This is evident from the posterior curves in Fig.\ref{fig:nuclboth}, which are colour-coded by the estimated virial halo masses for the systems under study. 
    For example, a $M_{I} = -15$ dwarf has an $\approx80\%$ probability of being nucleated in Coma, compared to $\approx50\%$ for quiescent satellites in groups with both early- and late-type centrals.
    In order to better quantify this dependence we introduce the $M_{I,f_{n50}}$ index, the magnitude at which the nucleation probability reaches 50\% as inferred from the logistic model. In Fig.\,\ref{fig:hostmass} we show how this half-nucleation magnitude varies as a function of the virial halo mass for each environment. We model this relation as a linear regression of the form $M_{I,fn50} = \alpha + \beta\, \text{log}(\mathcal{M}_{200})$. The mean posteriors and associated standard deviations obtained for the parameters are: $\alpha = -21.6\pm2.8$ and $\beta = 0.54\pm0.19$. This implies that the half-nucleation luminosity correlates with the host halo mass roughly as $L_{I,f_{n50}} \propto \mathcal{M}_{200}^{-0.2}$. The most extreme environments, Coma and the late-type groups, have $L_{I,f_{n50}}$ values that differ by almost a factor of ten in luminosity. A corollary to this result is that the nucleation fraction in Coma is the highest in all the studied environments over more than three decades in dwarf galaxy luminosity.

\end{itemize}

The main finding of this study is that \textit{the rate at which the probability of nucleation varies with magnitude does not depend substantially on the environment, but the fraction of nucleated dwarfs at a fixed luminosity does}. As a result, any model of NSC formation in low-luminosity/mass galaxies needs to account for the joint dependence of the nucleation fraction on galaxy luminosity/mass and the environment in which they reside.
We discuss possible scenarios in the next Section.

\section{Discussion}

\label{sec:discussion}

We now attempt to frame the two main results of this work in the context of NSC formation scenarios in low-luminosity/mass galaxies. The finding that the nucleation fraction displays a nearly-universal dependence with galaxy luminosity simply implies a higher ability of more massive dwarfs to pile up material in their central regions. 
Studies of NSC occupation alone cannot constrain whether this process occurs through star cluster inspiral or in situ star formation following gas inflows.
There is, however, mounting evidence that the former process is probably dominant for low-mass quiescent galaxies \citep{neumayer20}. 
Observations in support of this picture include the fact that NSCs in faint, early type dwarfs are typically more metal-poor than their host galaxies \citep{fahrion20,johnston20}; that the occupation fraction of GCs and NSCs track each other remarkably well \citep{sanchez-janssen19}; and that in this luminosity regime the scaling of NSC mass with galaxy mass is in excellent agreement with the prediction from GC inspiral models \citep{antonini13,gnedin14,sanchez-janssen19}.
Such models require  GCs (or their progenitors) to have masses high enough so that their orbits decay in less than a Hubble time as a consequence of dynamical friction, while simultaneously surviving tidal dissolution.
Factors that influence NSC growth and may depend on galaxy mass are the GC mass function (GCMF), the GC formation distance, the host galaxy structural properties, and the presence of a massive black hole--but the BH occupation fraction for low-mass galaxies is poorly constrained, and it certainly does not increase for fainter galaxies \citep{Greene20}.
Given that the nucleation fraction peaks at $M_I \approx -18$ ($\mathcal{M} \approx 10^9 \mathcal{M}_{\odot}$) and then declines, fainter dwarfs must have a comparatively bottom-heavier GCMF, or have GCs that on average form at larger distances from the galactic centre, or be preferentially concentrated and more compact.
There are no observational constraints on the first two properties, and the latter is in direct contradiction with the observed mass-size relation of faint dwarfs in the nearby Universe \citep{mcconnachie12,eigenthaler18,ferrarese20}.
Alternatively, it is possible that, even if the GCMF is universal, the formation of GCs in galaxies with such low masses remains a stochastic process.
If dwarfs were to form clusters so that the total mass of the GCS is a constant fraction of their stellar mass, this would explain why in the dwarf regime the NSC and GC occupation fractions track each other so closely: the nucleation fraction drops simply because the host galaxies never form enough massive GCs to begin with.
A corollary from this proposition is that the GCMF should vary with galaxy luminosity, with very faint dwarfs exhibiting a deficit in massive GCs. 

In the context of GC inspiral scenarios, the same arguments apply to the unambiguous environmental dependence we show in Figs. \ref{fig:nuclboth} and \ref{fig:hostmass}--but now the differences must arise due to effects related to the host halo at fixed present-day galaxy luminosity.
Unfortunately, to our best knowledge the environmental dependence does not arise naturally in any NSC formation model, and there are no observational constraints on the early properties of  GCSs in different environments.
Additionally, the mass-size relation seems to be rather universal in this luminosity regime, with dwarfs in environments ranging from Coma to the Local Group following similar scaling relations.
\citet{sanchez-janssen19} speculate that a biased formation scenario for star clusters similar to that proposed by \citet{peng08} does a reasonable job at explaining the environmental dependence of the faint nucleation fraction, at least qualitatively.
In such a model galaxies that now reside in higher density environments form stars earlier and sustain higher star formation rates (SFR) and SFR surface densities. 
These are conditions conducive to the formation of bound massive
clusters, and if cluster formation efficiency is close to universal and galaxies formed clusters proportionally to their mass at early epochs \citep{kruijssen15} then one naturally expects a larger mass fraction in star clusters in the more biased environments.
In this context it is also important to remember that, for a given present-day stellar mass, the subhaloes in denser environments were at all times prior to infall more massive than those in less dense regions.
This, together with earlier infall (and peak mass) times creates conditions that favour efficient formation of star clusters \citep{mistani16}. 
Those that are massive enough and form close to the galaxy centre will experience dynamical friction and decay to the bottom of the potential well.
We finally note that some of the masses of NSCs in very faint galaxies are comparable to those of typical GCs (see Fig.\,\ref{fig:nucgal}), and therefore very little GC merging is actually required \citep{fahrion20, neumayer20}. 

The proposition that nucleated dwarfs form a biased subpopulation is well established by numerous observational results. 
Compared to their non-nucleated counterparts they exhibit more concentrated spatial distributions and a propensity for circularised orbits \citep{ferguson89,lisker07,lisker09}; they are intrinsically more spherical \citep{ryden94,sanchez-janssen19b} and possess more concentrated light profiles \citepalias{denbrok14}; they host older stellar populations \citep{lisker08}; and there is tentative evidence that they feature higher GC mass fractions \citep{miller98,sanchez-janssen12}.
Whether or not this is sufficient to explain the higher occurrence of NSCs in more massive haloes remains to be quantified by detailed models of their formation and evolution. 
 
\section{Summary and Conclusions}
\label{sec:conclusion}

In this work we detect and characterise 66 low-mass, quiescent galaxies in the central regions of the Coma cluster using deep HST/ACS imaging in the F814W band. NSCs are identified by a combination of visual inspection and full image modelling. We perform Bayesian logistic regression to model the joint dependence of the nucleation fraction on galaxy absolute magnitude and environment for dwarf galaxies in nearby clusters and groups. Our main conclusions are:

\begin{itemize}
    \item[1)] Similar to previous findings \citep{denbrok14,sanchez-janssen19}, fainter galaxies in Coma tend to show a smaller difference in magnitude from their nuclei--but the significant scatter at fixed galaxy luminosity suggests the growth of stellar nuclei is a substantially stochastic process.
     \item[2)] The nucleation fraction depends on both galaxy mass/luminosity and environment, with the former being the primary parameter. Fainter galaxies have a lower probably of hosting NSCs, as do quiescent satellites of all luminosities in lower mass haloes. The rate at which the probability of nucleation varies with luminosity is remarkably universal.
    \item[3)] The nucleation fraction in Coma over three decades in dwarf galaxy luminosity is higher than in any other known environment. This is a direct result of the strong environmental dependence of the nucleation fraction. We find that the luminosity at which half of the dwarf galaxies contain an NSC is inversely proportional to the virial mass of the host halo, $L_{I,f_{n50}} \propto \mathcal{M}_{200}^{-0.2}$.

\end{itemize}

We have shown that nucleation in dwarf galaxies is a complex phenomenon that depends both on luminosity and the environment in which the galaxy resides. 
We identify several observational constraints that would advance our knowledge of the conditions and environments that are conducive to the formation and growth of stellar nuclei.
First, it is critical to rise the statistical significance of the environmental dependence by studying the nucleation fraction in other groups and clusters--and in particular for both more and less massive host haloes than studied here. 
The rarity of massive clusters like Coma and the sparseness of low-density environments make this a challenging task that might be best tackled with upcoming wide-field space missions like Euclid, the Roman Space Telescope and the Chinese Space Station Telescope.
Second, the availability of large NSC samples would allow us to investigate if other physical parameters such as galaxy size also play an important role on the presence of nuclei, as predicted by models \citep{mistani16, antonini15}. 
Finally, occupation studies have little constraining power on the exact physical processes that drive NSC formation and growth. 
Detailed chemo-dynamical studies of nuclei across a wide range of masses and environments will be instrumental in settling this question \citep{kacharov18,fahrion19,johnston20, carlsten21}.

\section*{Acknowledgements}

We thank the anonymous referee for all the insightful suggestions that greatly improved the presentation of this paper. We also thank R. Flores-Freitas for the insightful contributions. This work is based on observations with the NASA/ESA \textit{Hubble Space Telescope}, obtained at  the Space Telescope Science Institute, which is operated by AURA, Inc., under NASA contract NAS\,5-26555. These observations are associated with GO Programs \#11711. EZ acknowledges funding from \textit{Conselho Nacional de Desenvolvimento Cient\'ifico e Tecnol\'ogico} (CNPq) through grant CNPq-162480/2017-2, \textit{Coordenação de Aperfeiçoamento de Pessoal de Nível Superior} (CAPES) and the Newton Fund. ACS acknowledges funding from CNPq and \textit{Fundação de Amparo à Pesquisa do Estado do Rio Grande do Sul} (FAPERGS) through grants CNPq-403580/2016-1, CNPq-11153/2018-6, PqG/FAPERGS-17/2551-0001, FAPERGS/CAPES 19/2551-0000696-9 and L'Or\'eal UNESCO ABC \emph{Para Mulheres na Ci\^encia.} JPB was supported in part by the International Gemini Observatory, a program of NSF’s NOIRLab, managed by AURA, Inc., under a cooperative agreement with the National Science Foundation, on behalf of the Gemini partnership.
 
 \section*{Data availability}
 
 The data underlying this article are available in the article and at the MAST HST archive (https://archive.stsci.edu/hst/) under the program ID 11711.




\bibliographystyle{mnras}
\bibliography{ref.bib} 




\appendix
\section{References for the data used in this work not from the Coma cluster}
\label{appendixA}

In this work we study the nucleation fraction in the Coma cluster and in other environments using literature data. In Table\,\ref{tab:appendixA} we list the other systems included in the analysis, namely the Virgo and Fornax clusters, as well as Local Volume groups ($D < 12$ Mpc) with both early-type (ET) and late-type (LT) centrals. In the table we indicate the number of quiescent satellites in each environment, as well as the literature sources for the photometry and the nucleation classification. 

Where applicable, we have converted the published magnitudes to the $I$-band. We adopt conversions based on the ones presented in \citet{blanton07}:

\begin{align}
    (B-i) \approx 1.10, \\
    (i - I) \approx 0.06.
\label{eq:conversions}
\end{align}

The selection of quiescent satellites in the Local Group and around M81 is detailed in \citet{sanchez-janssen19}.
The majority of the low-density systems are drawn from \citetalias{carlsten20a}. We only select galaxies classified as dEs and discard dIrr and transition dwarfs. \citetalias{carlsten20b} present surface brightness fluctuation distances for the galaxies in the original sample. We make use of these and consider only galaxies that are flagged as "possible" or "confirmed" satellites in their table 4. The nucleation classification and photometry for the final sample of group satellites is then taken from \citetalias{carlsten20a} and \citetalias{carlsten20b}, respectively.  

\begin{table*}
\caption{Source of the photometry and nucleation classification for the data used in this work. From left to right: First column is the central galaxy for a given group in the local volume or the galaxy cluster. For Local Volume groups the second column indicates the morphology of the central galaxy (ET for early-types and LT for late-types). The next column is the number of quiescent satellite galaxies in each system, with the two subsequent columns presenting the sources for the photometry and the nucleation classification, respectively.}
\label{tab:appendixA}
\resizebox{\textwidth}{!}{%
\begin{minipage}{\textwidth}
\begin{tabular}{lllll}
\hline
\multicolumn{2}{l}{}        & \multicolumn{3}{c}{Local Volume Groups}                                                \\ \hline
ID          & Morph.     & N   &  Photometry Source               &  Nucleation  Source                    \\ \hline
NGC 1023       & ET          & 15  & \citet{carlsten20b}              & \citet{carlsten20a}                    \\
M104           & ET          & 23  & \citet{carlsten20b}              & \citet{carlsten20a}                    \\
Cen A & ET & 31 & \citet{carlsten20b},\citet{muller19}\footnote{\citet{muller19} is the source photometry for the galaxies KK54 and KK58, and source nucleation for KK54.} & \begin{tabular}[c]{@{}l@{}}\citet{carlsten20a},\\ \citet{muller19}, \citet{fahrion20}\footnote{\citet{fahrion20} is the source nucleation for the galaxy KK58.}\end{tabular} \\
NGC 3115       & ET          & 24  &  Chies-Santos et al. (in prep.) &  Chies-Santos et al. (in prep.)      \\
NGC 4631       & LT          & 7   & \citet{carlsten20b}              & \citet{carlsten20a}                    \\
NGC 4565       & LT          & 16  & \citet{carlsten20b}              & \citet{carlsten20a}                    \\
NGC 4258       & LT          & 9   & \citet{carlsten20b}              & \citet{carlsten20a}                    \\
M51            & LT          & 4   & \citet{carlsten20b}              & \citet{carlsten20a}                    \\
MW             & LT          & 11  & \citet{karachentsev13}           & \citet{sanchez-janssen19}                 \\
M81            & LT          & 13  & \citet{karachentsev13}           & \citet{sanchez-janssen19}                 \\
M31            & LT          & 30  & \citet{karachentsev13}           & \citet{sanchez-janssen19}                 \\ \hline
\multicolumn{2}{l}{}        & \multicolumn{3}{c}{Galaxy Clusters}                                             \\ \hline
\multicolumn{2}{l}{ID} & N   &  Photometry Source              & \multicolumn{1}{l|}{Nucleation Source} \\ \hline
\multicolumn{2}{l}{Coma}    & 255 & This work, \citet{denbrok14}     & This work, \citet{denbrok14}           \\
\multicolumn{2}{l}{Virgo}   & 382 & \citet{sanchez-janssen19}         & \citet{sanchez-janssen19}               \\
\multicolumn{2}{l}{Fornax}  & 263 & \citet{munoz15}                  & \citet{munoz15}                       
\end{tabular}%
\end{minipage}
}
\end{table*}

\section{Additional details from the hierarchical Bayesian logistic analysis}
\label{appendixB}

In this section we discuss briefly some additional details from the Bayesian analysis described in Section \ref{sec:logit}. 

\begin{figure}
    \centering
    \includegraphics[width=\linewidth]{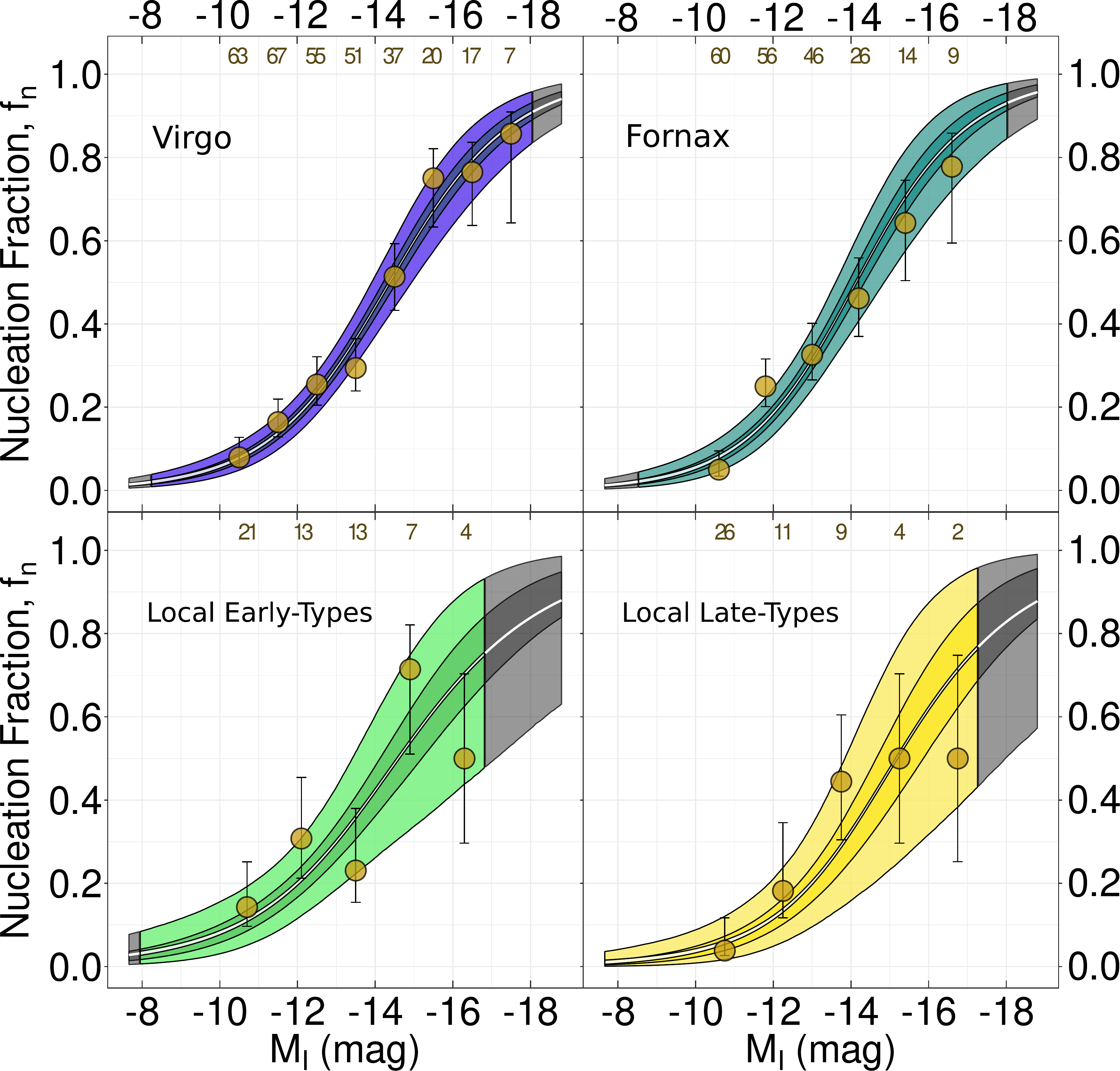}
    \caption{Same as the bottom panel of Fig. \ref{fig:nucl}, but for every other environment considered in this work, gathered from the literature references described in section \ref{sec:other}. Notice the excellent agreement between the binned data and the logistic regression.}
    \label{fig:individual}
\end{figure}

In Fig. \ref{fig:individual} we show the individual posterior distributions from the Bayesian logistic regression alongside binned data points for every environment from the literature considered in Fig. \ref{fig:nuclboth}, with the same captions as in the bottom panel of Fig. \ref{fig:nucl}. It is clear that the logistic regression follows closely the binned data despite not relying on arbitrary defined bins. We stress the advantages of not relying on a binned fit, specially when dealing with data from different sources and with different sample sizes. Notice that the local environments have noticeable larger confidence intervals than Virgo, Fornax and Coma (show in Fig. \ref{fig:nucl}). Our method of choice is able to efficiently compare the nucleation fraction in different environments in a homogeneous way while not underestimating the uncertainties arisen from small sample sizes. 

\section{Estimation of photometric uncertainties}
\label{appendixC}

To estimate the uncertainties in the recovered parameters shown in Table \ref{tab:catalog} we run the photometry procedure presented in Section \ref{sec:phot} for mock galaxies randomly placed in the HST images used in this work. Mock galaxies are created using \textsc{galfit} using a S\'ersic $+$ PSF component, with parameters sampled from the values in Table \ref{tab:catalog}. Galaxy magnitudes are chosen randomly from an uniform distribution $U_{M_{gal}}(-14.0, -10.0)$. NSC magnitudes, effective radii and S\'ersic indices show significant correlations with galaxy magnitude, with Pearson correlation coefficients of 0.42, -0.54 and -0.44 respectively. Due to this, we group these quantities in 10 bins of galaxy magnitude and calculate the mean, $\mu_{bin}$, and standard deviation $\sigma_{bin}$. Then, in each bin, we randomly generate new values for such quantities following a normal distribution $N(\mu_{bin},\sigma_{bin})$. We do not detect a significant correlation between axis ratios and galaxy magnitude (Pearson correlation coefficient of -0.09). For this reason, mock values for the axis ratio are obtained from an uniform distribution $U_{q}(0.35, 1.0)$. Position angles are set to a fixed value of 75 degrees for every mock galaxy, as this parameter has little impact on the uncertainties we aim to estimate. This sampling technique ensures that we simulated only realistic representations of galaxies as detected in our catalogue. We created a total of 10,000 nucleated galaxies, 5,000 for each HST field. Mock galaxies are added in random positions in the HST images with all galaxies masked. Then we proceed with photometry exactly as done for our real detections and described in Section \ref{sec:phot}, except for the step of visual classification.   

\begin{figure}
    \centering
    \includegraphics[width=\linewidth]{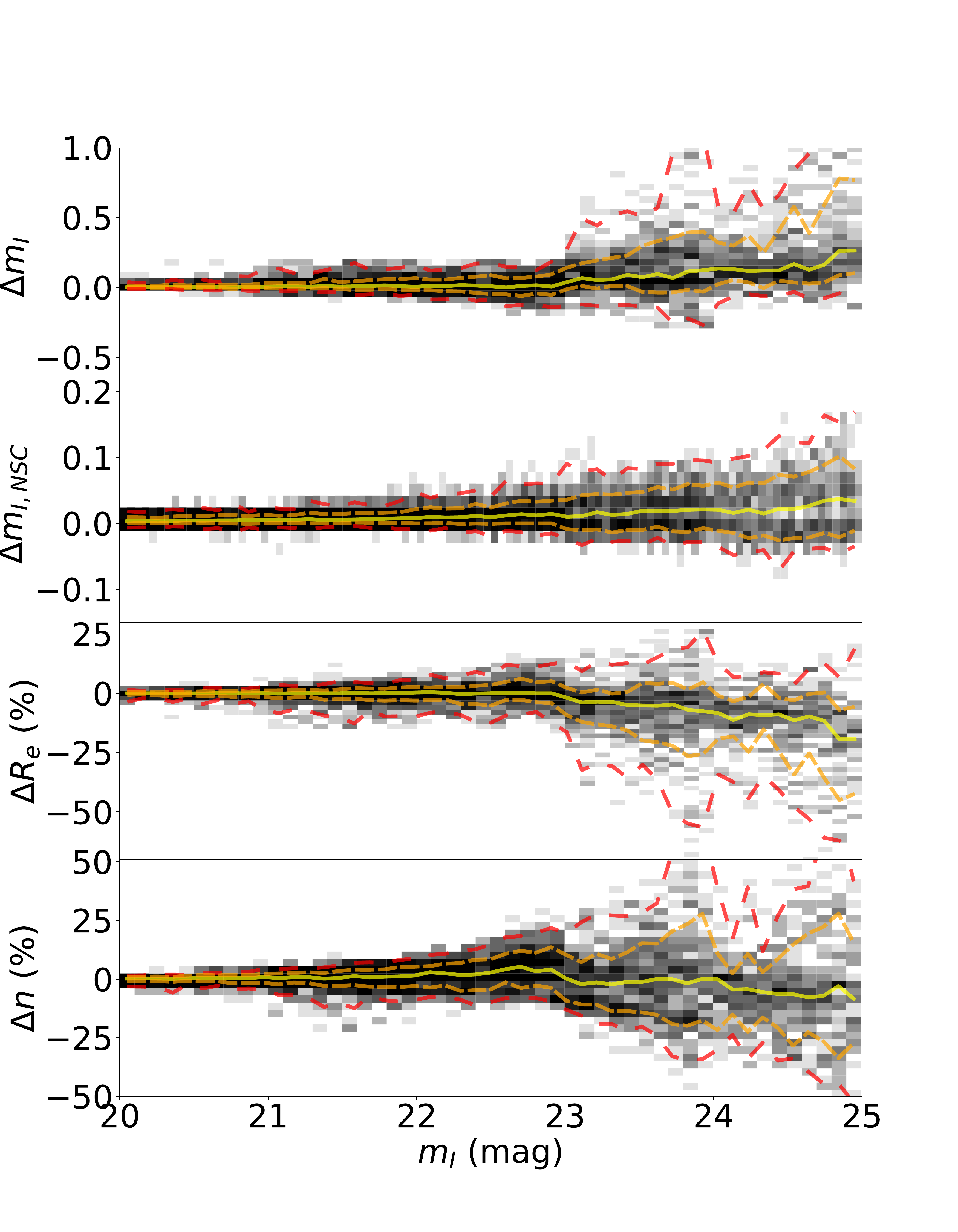}
    \caption{Results of our photometry procedure applied to 10,000 mock galaxies randomly positioned in both HST/ACS fields studied in this work; 5,000 in the NGC 4874 field and 5,000 in the NGC 4889 field. We show the difference between the "true" and "fitted" quantities in the form $\Delta$X, where X can be galaxy magnitudes ($m_{I, gal}$), NSC magnitudes ($m_{I,NSC}$), effective radii ($R_e$) and S\'ersic index ($n$). For $R_e$ and $n$ we show this difference in percentages. In the x-axis we show "true" values of $m_{I, gal}$. The grey scale shows the density of galaxies in bins of galaxy magnitude. Solid yellow lines show the median distribution of points, while orange and red dashed lines show the 95\% and 68\% confidence intervals, respectively.}
    \label{fig:sim}
\end{figure}

The results of these simulations are presented in Fig. \ref{fig:sim}. The median uncertainty in galaxy magnitude is $\sim$ 0.2 mag. The 95\% and 68\% confidence intervals show that \textsc{galfit} estimates magnitudes brighter than the true values for galaxies fainter than $M_{I}$ $\sim$ 11 mag. This behaviour was also observed in the simulations of \citet{ferrarese20} for the NGVS. Therefore, we do not consider this to be evidence for any systematic bias from our method or observations. Regarding NSC magnitudes the uncertainties are considerably lower, never surpassing 0.1 mag even considering the 95\% confidence intervals. Finally, results for effective radii and S\'ersic index show a median uncertainty of $\sim$ 20\%. 

\bsp	
\label{lastpage}
\end{document}